\documentclass[preprint,superscriptaddress]{revtex4-1} 
\usepackage{hyperref} 
\usepackage{amssymb}
\usepackage{graphicx}

\begin{document}

\title{Optically optimal wavelength-scale patterned ITO/ZnO composite coatings for thin film solar cells}

\author{Antoine Moreau}
\affiliation{Center for Metamaterials and Integrated Plasmonics, Duke University, Durham, North Carolina 27708, USA}
\affiliation{Clermont Universit\'e, Universit\'e Blaise Pascal, Institut Pascal, BP 10448, F-63000 Clermont-Ferrand}
\affiliation{CNRS, UMR 6602, IP, F-63171 Aubi\`ere}
\author{Rafik Smaali}
\author{Emmanuel Centeno}
\affiliation{Clermont Universit\'e, Universit\'e Blaise Pascal, Institut Pascal, BP 10448, F-63000 Clermont-Ferrand}
\affiliation{CNRS, UMR 6602, IP, F-63171 Aubi\`ere}
\author{Christian Seassal}
\affiliation{Universit\'e de Lyon, Institut des Nanotechnologies de Lyon (INL), UMR 5270 CNRS-INSA-ECL-UCBL}
\affiliation{Ecole Centrale de Lyon, 36 avenue Guy de Collongue, 69134 Ecully Cedex, France}

\begin{abstract}
A methodology is proposed for finding structures that are, optically speaking, locally optimal : a physical
analysis of much simpler structures is used to constrain the optimization process. The obtained designs are
based on a flat amorphous silicon layer (to minimize recombination) with a patterned anti-reflective coating 
made of ITO or ZnO, or a composite ITO/ZnO coating. These latter structures are realistic and present 
good performances despite very thin active layers.
\end{abstract}

\maketitle

\section{Introduction}

Thanks to the maturity of nanophotonics, various advanced light
trapping schemes suited to photovoltaic (PV) solar cells have been
proposed and investigated in the recent years. The interest of these
novel approaches is to generate highly efficient anti-reflecting
structures\cite{haase07,lo07,kroll08,dewan09,campa10,zanotto10,mad11}, but also, and maybe more importantly,
to increase significantly the absorption efficiency in the long
wavelength range - where active media are generally less efficient - and
finally to reduce the thickness of the absorbing layers.  Different
groups have been focusing on the possibility to pattern the absorbing
layers like crystalline or amorphous silicon\cite{park09,eldaif10,mallick10,seassal10}. 
This leads to a high absorption enhancement, but the positive impact on the conversion
efficiency is still to be assessed. Indeed, the increased area
corresponding to the free and processed surfaces may lead to extensive
surface recombination. Moreover, the designs generally result from
simple design rules, or geometrical parameters scanning. There is
therefore a strong need to develop more advanced methodologies, in
order to optimize the design of these advanced solar cells.  In this
paper, our goal is (i) to propose a design methodology based on a 
purely numerical optimization, guided by a physical analysis, that
would be able to make locally optimal structures emerge from the process
(ii) to provide designs for solar cell structures
including an anti-reflective layer patterned as a photonic lattice,
and compatible with a low level of surface recombinations, using 
the previous methodology.

Here we will consider both flat and periodically patterned anti-reflective coatings
made of ITO and ZnO, the two most common
materials used as transparent conducting electrodes. We carry on a
thorough study of the simplest structures and use genetic algorithms
guided by our physical analysis to optimize grating structures.
This will lead us to propose hybrid anti-reflective coatings made of ITO and ZnO,
to combine the optical properties of ZnO (which exhibits a high refractive 
index and a low absorption coefficient\cite{liu06}) with the electrical 
properties of ITO (which is more conductive, but also more absorbent). The
structures we propose are finally very thin (typically thinner
than $300\,\mbox{nm}$) but present surprisingly good efficiencies. These
structures can thus be considered as an extremely efficient way to
use silicon to convert light into electron-hole pairs and may be 
relatively inexpensive to produce.

\section{Flat anti-reflective coatings}

The goal of this first part is to look for very simple but optically optimal
structures and to understand the physical reasons why these structures 
are locally optimal. This will allow us to guide the numerical optimization
of the more complex structures that will be studied in section \ref{deux}.

\begin{figure}[h]
\begin{center}
\includegraphics[width=8cm]{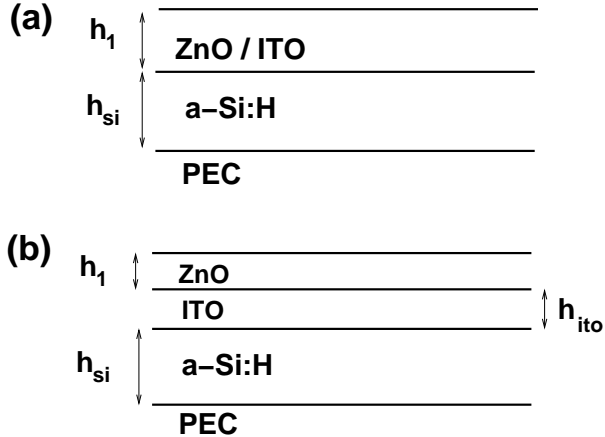}
\end{center}
\caption{(a) Layer of amorphous silicon (of thickness $h_{si}$) backed
  with a perfect electric conductor (PEC) with a simple layer of ITO or ZnO
  (thickness $h_1$) as an anti-reflective coating and (b) Hybrid coating with a
  thickness for the ITO layer of $30\,\mbox{nm}$.\label{schema1}}
\end{figure}

In this first part, we will consider flat anti-reflective coatings made
of ZnO or ITO (as shown in figure \ref{schema1} (a)), or bi-layers {\em hybrid} 
coatings with an ITO layer for electrical reasons and a ZnO layer on top to 
improve the optical properties of the coating (see figure \ref{schema1} (b)).
A thickness of $30$ nm has been considered for the ITO layer in the
case of a hybrid coating. Reducing the ITO thickness down to 30nm, compared to a more standard 
value around 50-70 nm\cite{haug11}, yields a lower useless absorption in the TCO layer. 
This reduction is made possible in the case of our hybrid design, since 
the top ZnO layer will also contributes to the lateral conduction of 
electrical charges.

The a-Si:H layer stands above a perfect electric conductor for
simplicity, and to be able to compare with previous works on this 
subject\cite{zanotto10}, some of them being particularly reliable and
thorough\cite{kroll08}.

The figure of merit in all our work is the short-circuit current under
normal illumination. We assume that all the photons absorbed by the
silicon are converted into electron-hole pairs and that all the
carriers produced contribute to the current - the recombinations of
the electron-hole pairs being minimized when considering a thin
enough, flat a-Si:H layer. The short-circuit current can in that case be
written
\begin{equation}\label{eq:jsc}
j_{sc} = \int A(\lambda)
\frac{dI}{d\lambda}. \frac{e\lambda}{hc}\,\mbox{d}\lambda
\end{equation}
where $\lambda$ is the wavelength, ranging from $375$ nm (beginning
of the solar spectrum) to $750\,\mbox{nm}$ (above which the a-Si:H becomes
transparent), $\frac{dI}{d\lambda}$ the spectral energy density of the
light source (we have taken an am1.5 normalized spectrum\cite{am1.5}) and
$A(\lambda)$ the absorption of the a-Si:H layer (see Appendix \ref{appendix} for details
regarding the numerical computation of the absorption). If all the incoming photons of the
considered spectral range were converted into electron-hole pairs, the
short-circuit current would be equal to $j_0=23.665\,\mbox{mA}/\mbox{cm}^2$. The
{\em conversion efficiency} ($CE$) is given by the fraction of the
incoming photons that are converted into electron-hole pairs
i.e. $CE=\frac{j_{sc}}{j_0}$.

\begin{figure}[h]
\begin{center}
\includegraphics[width=8cm]{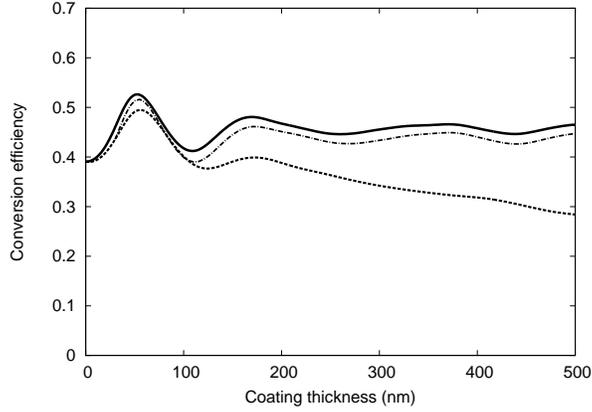}
\end{center}
\caption{Conversion efficiency as a function of the thickness of an
  anti-reflective coating made of ZnO (solid line), ITO (dashed
  line) or hybrid anti-reflective coating of ZnO on top of a $30\,\mbox{nm}$ thick ITO
  layer (dashed-dotted line). The thickness is the overall thickness of the
  coating in the hybrid case.}
\label{fig1}
\end{figure}

Figure \ref{fig1} shows the conversion
efficiency of the structure when either ZnO or ITO anti-reflective coatings (on top
of a 100 nm thick a-Si:H layer) are considered, as a function of the
coating thickness $h_1$. It appears clearly that there is an obvious
optimal thickness around 52 nm for ZnO (and 56 nm for ITO) for a
$100\,\mbox{nm}$ thick silicon layer. This optimal thickness for the coating
is accurate : the efficiency is roughly 20\% smaller for a 100 nm anti-reflective
coating in both cases. This behavior is universal 
and does not depend on the thickness of the amorphous silicon layer, even 
if the {\em optimal} thickness slightly depends on the a-Si:H thickness.

For thick ZnO layers, the absorption tends to a limit that is $10\%$
below the maximum (this difference is even worse for thicker silicon layers).
For ITO the absorption by the silicon layer is a decreasing function of the thickness due to the
absorption by the coating itself. This confirms that thin and higher index anti-reflective coatings 
are a better solution from an optical point of view\cite{zanotto10}, as long as the losses 
can be neglected. But ZnO does not
present a very high conductivity, so that it is usually necessary to use very thick
ZnO layers\cite{kroll08}. That is obviously detrimental to the optical properties,
hence the idea to combine a thin ITO layer with a ZnO layer. The resulting anti-reflective coating
presents intermediary optical properties, as shown in figure \ref{fig1}, making this structure
a very interesting trade-off.

\begin{figure}[h!]
\begin{center}
\includegraphics[width=8cm]{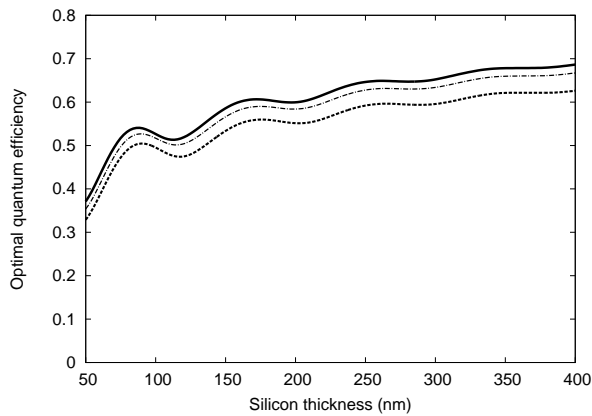}
\end{center}
\caption{Conversion efficiency of the structure as a function of the
  silicon thickness for an optimal coating made of ZnO (solid line),
  ITO (dashed line) or hybrid anti-reflective coating of ZnO on top of a 
  $30\,\mbox{nm}$ thick ITO layer (dashed-dotted line).}
\label{fig2}
\end{figure}

Let us now study how the absorption could be influenced by the
thickness of the a-Si:H layer itself, when an optimal anti-reflective coating is
assumed ({i.e.} for each thickness of the a-Si:H layer, the optimal
thickness of the anti-reflective coating is computed and used to find the CE). The
results are shown  in figure \ref{fig2}.  The absorption is not a strictly
increasing function of the silicon layer thickness, as long as this
thickness is smaller than 300 nm.  This means that there are clear
local maxima of the CE, and that this thickness should probably not
be chosen arbitrarily when it is small.

Figure \ref{fig2} shows three local maxima, whatever the anti-reflective coating. The first two maxima are more
pronounced than the third one, and since they happen for small values of the thickness, these
maxima correspond to structures making a much more efficient use of silicon - that is especially
true for the first maximum, but the second one definitely present a higher efficiency that
could be interesting too. The characteristics of these structures are detailed table \ref{table1}.

\begin{table}[h]
\begin{center}
\begin{tabular}{ l | c  c  | c  c }
Structure & $h_{si} \,(nm)$ & $h_1\,(nm)$ & CE & $j_{sc} \,(\mbox{mA}/\mbox{cm}^2)$\\ 
\hline \hline 
ITO, 1st max.  & 90.3 & 54.6 & 50.2 & 11.9 \\
ZnO, 1st max.  & 87 & 51 & 54.2 & 12.8 \\
Hybrid, 1st max. & 89.8 & 22.7 & 53 & 12.5 \\
ITO, 2nd max. & 175.8 & 57.5 & 55.75 & 13.2 \\
ZnO, 2nd max.  & 172.1& 54.2 & 60.6 & 14.3 \\
Hybrid, 2nd max. & 174.7& 25.86 & 59.0 & 14 \\
\end{tabular}
\end{center}
\caption{Geometrical parameters of the local optical optima shown in figure \ref{fig2}.\label{table1}}
\end{table}

\begin{figure}[h]
\begin{center}
\includegraphics[width=8cm]{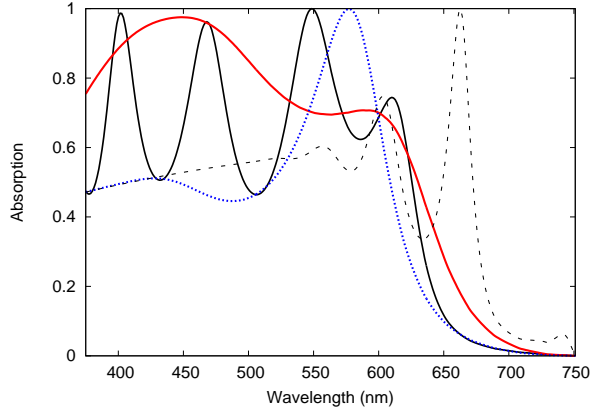}
\end{center}
\caption{Absorption spectra for different structures : a $87\,\mbox{nm}$ thick 
a-Si:H layer (i) with a $51\,\mbox{nm}$ thick optimal ZnO coating (red curve)
(ii) with a $500\,\mbox{nm}$ thick ZnO coating (solid black curve) (iii) without
any anti-reflective coating (blue dotted curve) and (iv) a $500\,\mbox{nm}$ thick a-Si:H layer
without coating (black dashed curve).\label{spectres_flat}}
\end{figure}

Physically, the efficiency of an anti-reflective coating relies on the resonances
it may support : any resonance is associated with a high transmission
of light to the underlying medium. If the anti-reflective coating presents an index
that is intermediate between the active layer and the superstrate (the superstrate being air, this is the case for ITO and ZnO here), 
these resonances occur when $\lambda \simeq \frac{4h\,n}{1+2\,m}$ where $h$ is the thickness 
and $n$ the index of the dielectric layer and $m$ is the order of the resonance.
Figure \ref{spectres_flat} shows absorption spectra for the optimal thickness
of a ZnO layer and for a much thicker layer. For thick layers, the
resonances are numerous and narrow. Slightly changing the thickness in this
regime will just shift the resonances a little, but since the resonances 
and the anti-resonances will be averaged over the whole spectrum, 
the change in the thickness has little impact on the overall efficiency.
On the contrary, a very thin anti-reflective coating supports essentially one resonance
around $\lambda \simeq 4 h\,n(\lambda)$. The position of this {\em
  broad} resonance has a great impact on the CE and it can be tuned to
maximize the efficiency : the optimal position for the resonance seems 
to be located where the silicon is the most absorbent. According to
the formula above, a thickness of $51\,\mbox{nm}$ should produce a resonance
around $435\,\mbox{nm}$ - which is totally consistent with the spectra shown in figure 
\ref{spectres_flat}. That is why the efficiency is very sensitive to the
thickness of the coating for thin layers, and why thin coatings are more
efficient\cite{zanotto10}.

Let us now consider the impact of the silicon layer on the CE.
The a-Si:H layer can support cavity resonances when the penetration length is
smaller than the thickness. A typical absorption spectrum for a thick layer
without any anti-reflective coating is shown in figure \ref{spectres_flat}. 
For short wavelength ($\lambda<550\,\mbox{nm}$), the amorphous silicon is very absorbent
and the spectrum is almost flat. Several resonances can be seen on the spectrum
for longer wavelength. As discussed already for the anti-reflective coating, these resonances are narrow because
of the layer's thickness. Any change in the thickness is likely to increase 
the conversion efficiency, but only slightly. On the contrary, thin a-Si:H
layers (compared to the penetration length) may support broad resonances over
almost the whole spectrum. As shown in figure \ref{spectres_flat}, a $87\,\mbox{nm}$ thick
silicon layer without any coating supports essentially one broad resonance in the
middle of the visible spectrum. This resonance still appears as a shoulder on the
spectrum of the structure with the optimal anti-reflective coating. 

Finally, these carefully optimized cells present good theoretical efficiencies. 

Let us consider a typical non-optimized flat structure, also considered in \cite{kroll08}.
The structure is a $300\,\mbox{nm}$ thick a-Si:H layer, backed by a $19\,\mbox{nm}$ ZnO layer and a 
perfect electric conductor, with a $1024\,\mbox{nm}$ thick ZnO coating. According to our simulations,
it presents an efficiency of $53.3\%$ on the range of the spectrum considered here (our simulations
are generally in an excellent agreement with all the results presented in Kroll {\em et al.}). This
structure is typical essentially because, for electrical reasons, it presents a very thick 
transparent oxide layer.

By comparison, the locally optimal structures we propose are able to reach a similar efficiency with much 
less silicon (three times less silicon typically in this case). This is of course partly due to the optimized anti-reflective coating, 
and can be seen as an illustration that, as shown in figure \ref{fig2}, it is relatively easy to reach interesting efficiencies with thin active layers,
while any further increase of the efficiency may generally require a large supplementary amount of silicon.

\section{Patterned coatings\label{deux}}

\begin{figure}[h]
\begin{center}
\includegraphics[width=8cm]{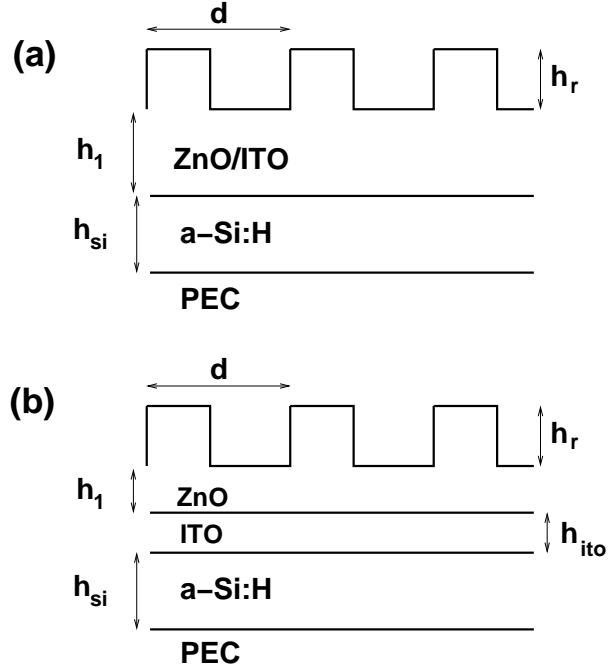}
\end{center}
\caption{Structured anti-reflective coating. The geometrical parameters 
  are $h_{si}$, the thickness of the silicon layer, $d$ the period of the grating, $h_r$ its height
  and $f$ the fill factor, while $h_1$ is the thickness of the layer
  between the silicon and the grating.\label{schema2}}
\end{figure}

It is well known that structuring the silicon or the electrode
layer can have a great impact on absorption, leading to increased
efficiencies. In this part, the impact of the patterning of the anti-reflective coating
only on the conversion efficiency is assessed. Our goal is still to
find optically (and at least locally) optimal structures with a 
flat and as thin as possible amorphous silicon layer, to minimize
recombination. The structures we consider are shown in figure \ref{schema2}.

The optical behavior of these structures is complex, and their geometry is
controled by many parameters.  The optimization cannot be done manually 
as it has been done above. Home-made genetic algorithms\cite{goldberg89} were thus used
to find parameters corresponding to a maximum conversion
efficiency.

Had we just ran an unconstrained optimization, the result would have been a very thick
active layer. But the previous physical analysis of the flat structures gives a strong indication
that locally optimal structures may exist even for more complex structures. In order to find these
local optima, the genetic algorithm has been constrained. Since we expected the optimal patterned structures to resemble somehow 
the flat ones, we have used different constraints based on the above results. The algorithm was allowed to consider only structures with 
$h_{si}< 150\, nm$, or with $h_{si}< 200\, nm$ (or even with $h_{si}< 250\, nm$).

The algorithm was able to find optimal structures, thus establishing the existence of local optima. The
results are summarized in table\ref{table2}. They show that the patterned optimal structures are very close
to the flat ones : the thickness of the active layer is very close in both cases, as is the thickness of
the homogeneous part of the anti-reflective coating. This is what allows us to conclude that the physical behavior of 
the structures is still dominated by the phenomena studied in the previous part : the resonances inside
the silicon layer and the homogeneous (unpatterned) part of the anti-reflective coating.

\begin{table}
\begin{center}
\begin{tabular}{ l | c  c  c  c  c | c  c }
Structure & $h_{si} \,(nm)$ & $h_r\,(nm)$ & f & $d \,(nm)$ & $h_1\,(nm)$ & CE & $j_{sc} \,(\mbox{mA}/\mbox{cm}^2)$\\ 
\hline \hline
ITO, $h_{si}<150\,\mbox{nm}$ & 91.6 & 39.35 & 0.355 & 434.4 & 44.0 & 50.96 & 12.06 \\
ZnO, $h_{si}<150\,\mbox{nm}$ & 94 & 207 & 0.35 & 439 & 53.5 & 61.3 & 14.5 \\
Hybrid,$h_{si}<150\,\mbox{nm}$ & 96.5 & 205 & 0.37 & 445 & 24.1 & 58.15 & 13.8\\
ITO, $h_{si}<200\,\mbox{nm}$ & 174.3 & 25.1 & 0.633 & 385.9 & 43.2 & 56.2 & 13.3 \\
ZnO, $h_{si}<200\,\mbox{nm}$ & 187 & 275 & 0.28 & 493 & 51.7 & 64.8 & 15.35 \\
Hybrid, $h_{si}<200\,\mbox{nm}$ & 188.4 & 370 & 0.62 & 415.7 & 38.3 & 62.2 & 14.7\\
ZnO, $h_{si}<350\,\mbox{nm}$ & 268 & 208 & 0.4 & 474 & 57.4 & 68.2 & 16.14 \\
\hline
\end{tabular}
\end{center}
\caption{Short-circuit current and conversion efficiency (CE) for various optimized structures depicted in figure \ref{schema2}.\label{table2}}
\end{table}

Moreover, let us just underline that :
\begin{itemize}
\item The patterning is obviously not interesting for the ITO coating
  : any increase of the ITO thickness is detrimental to the optical
  properties and the advantages of the patterning are not large enough
  to overcome that drawback - so that the flat structure seems to be
  very close to the optimal solution in that case.
\item The ZnO coating presents once more the best optical
  properties. It is possible to compare similar patterned and
  non-patterned structures. The patterning brings a increase of $13\%$
  in the conversion efficiency for the thinnest structure and of $7\%$
  for the second maximum.
\item The composite coating allows once more to approach the
  performances of the ZnO coatings. The patterning is responsible for
  a $10\%$ (respectively $6.2\%$) increase of the conversion efficiency for
  the first optimum (respectively the second optimum).
\end{itemize}

Physically, two mechanisms can explain the efficiency of the grating :
(i) it is a supplementary layer with a different effective
index, making the anti-reflective coating more effective for different wavelength,
(ii) it allows to excite guided modes inside the active layer,
leading to an enhanced absorption at longer wavelength where
the amorphous silicon can be considered almost completely
transparent. 

\begin{figure}[h]
\begin{center}
\includegraphics[width=8cm]{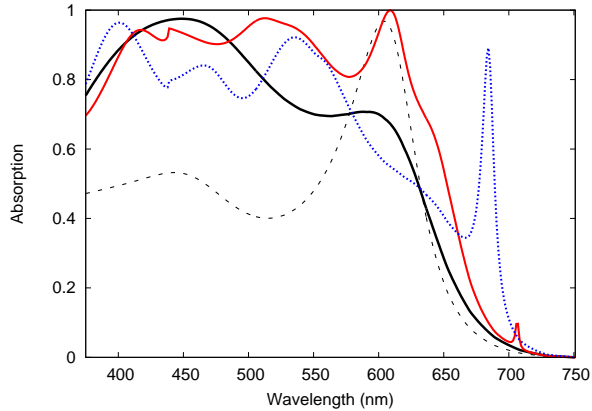}
\end{center}
\caption{Absorption of the first optimum for a ZnO coating (black
  line), of a $94\,\mbox{nm}$ a-Si:H slab (the black dashed line corresponds to the bare silicon layer) and of the first
  optimal grating structure (ZnO, $h_{si}<150\,\mbox{nm}$ ; red line : TE
  polarization, blue line : TM polarization).}
\label{spectres}
\end{figure}

Figure \ref{spectres} shows the absorption spectrum for
the optimal structures with a ZnO coating. The red curves on figure
\ref{spectres_flat} (optimal flat ZnO layer) and figure \ref{spectres} 
(optimal patterned ZnO coating) can be compared. The anti-reflective behavior persists
in the same way in the blue part of the spectrum. The most obvious difference
is the peak that can be seen around 600 nm. This can perfectly be understood
if the whole ZnO structure is seen as a double layer optical filter. The 
numerical method we use (see Appendix and \cite{moreau07}) gives access to 
the modes propagating inside a ZnO grating layer (corresponding to the first ZnO structure
in table \ref{table2}). Two modes are actually propagative : the first 
one has an effective index of 1.687 at 600 nm, and the second one presents an
lower index of 0.1149. The first mode can be considered responsible for
the transmission of this layer, the other being slightly excited but the incident plane wave. 
Adding the grating layer can thus be seen as adding another homogeneous layer to an optical filter, 
with a well defined refractive index and a simple simulation\cite{openfilters} shows
that the resulting anti-reflective coating presents a transmission peak at 600 nm in TE 
polarization. This perfectly explains the absorption peak of the structure.
There is obviously no absorption peak for the TM polarization, which
suggests that a 2D structure could be more effective if it was able to present
the same behavior for both polarizations.

\begin{figure}[h]
\begin{center}
\includegraphics[width=8cm]{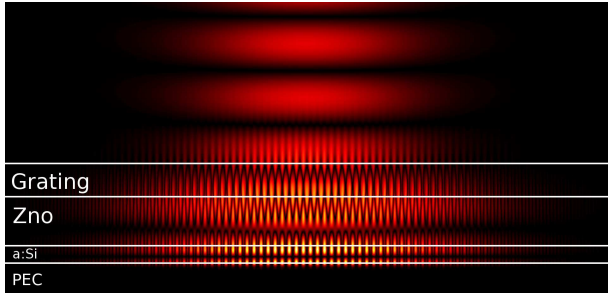}
\end{center}
\caption{Resonance excited in the structure corresponding to the first
  maximum for a ZnO anti-reflective coating structured layer, at $684.4\,\mbox{nm}$ but
  for a thickness $h_1=300\,\mbox{nm}$ : this allows to lessen the coupling 
 between the forward and the backward mode, allowing us to see the spatial
 extension due to the guided modes better.}
\label{resonance}
\end{figure}

The narrow peaks appearing in the red part of the spectrum are due to
guided modes inside the silicon layer excited by the grating. These
modes are actually leaky\cite{moreau08}. Figure \ref{resonance} shows
one of these particular resonances, whose position is dominated by the grating 
period and only marginally affected by the coating thickness. The fact that the field 
inside the silicon layer extends far from the excitation when 
illuminating the structure with a beam is a sign that a guided mode 
is actually excited. The extension of this mode is limited by the 
contra-directional coupling due to the grating - so that this phenomenon
looks very much like a light wheel here\cite{tichit07}.

These performances can be compared either to similar optimized structures\cite{kroll08}, 
to cells based on a thin (around $100\,\mbox{nm}$ thick) patterned amorphous 
silicon layer\cite{eldaif10,park09,campa10,zanotto10}. It is not always easy to make
accurate comparisons (especially when we disagree about the maximum achievable 
short-circuit current $j_0$\cite{zanotto10}). It still appears that the performances of the 
structures we have proposed are really good for so thin active layers.
For instance, with an optimized cell very similar to our, Kroll {\em et al.}
reach a $14.4\,\mbox{mA}/\mbox{cm}^2$ (according to our simulations, for the spectral 
range considered here) with a thick anti-reflective coating and a $338\,\mbox{nm}$ thick
silicon layer. According to our results, this performance could have been obtained 
with a $94\,\mbox{nm}$ thick layer and a very thin ZnO layer instead  - 
but to be fair, we should compare to the composite structure (the thin ZnO layer would not
be electrically realistic), that presents a $13.8\,\mbox{mA}/\mbox{cm}^2$ short-circuit 
current only. And the second optimum is based on a $188\,\mbox{nm}$ thick active
layer - but it presents slightly better properties ($14.7\,\mbox{mA}/\mbox{cm}^2$). Otherwise,
the absorption spectra of our structures compare very well with cells in which
recombinations are much more likely to happen\cite{eldaif10,park09,campa10},
reaching similar performances. 
Finally, Zanotto {\em et al.} see a $12.4\%$ increase of $j_{sc}$ due to a patterning
of the active layer that we were able to attain by patterning the anti-reflective coating only.

\section{Conclusion}

The thorough physical analysis we have conducted here shows
that the resonances occurring in very thin solar cells (with a flat active 
layer of a-Si:H typically thinner than $200\,\mbox{nm}$ and anti-reflective coatings based 
on a roughly $50\,\mbox{nm}$ thick homogeneous layer of ZnO or ITO)
have more impact on the conversion efficiency than for much thicker
structures. For these very thin structures, the short-circuit current
is for instance not a strictly increasing function of the silicon layer.
This means that there are pronounced local maxima of the conversion
efficiency for flat anti-reflective coatings as well as patterned anti-reflective coatings (increasing
the efficiency of at least $10\%$ for our thinnest structures compared to
flat coatings). We have used a genetic algorithm guided by
our physical analysis to find these locally optimal structures.

This kind of structures, with a flat and very thin active layer, is particularly interesting 
because (i) the recombinations are minimized, conversely to approaches based on
the patterning of the active layer and (ii) it represents a very efficient use
of the silicon to convert light into electron-hole pairs. We have actually obtained a
short-circuit current of $13.8 \mbox{mA}/\mbox{cm}^2$ for a structure with a $96.5\,\mbox{nm}$ thick 
only silicon layer (corresponding to a $58.15\%$ conversion efficiency over the 
$375$ to $750\,\mbox{nm}$ range). This structure has a composite anti-reflective coating made of a $30\,\mbox{nm}$ thick ITO
layer with a patterned ZnO layer on top of it, which allows to combine the optical properties
(high index and low losses) of ZnO with the electrical properties of ITO. 

Using the same methodology, more complex photonic patterns can be
considered, for example leading to a reduced dependence on the
incident light polarization or to an absorption enhancement over a
wider wavelength range. This includes 2D patterns or multi-scale
structures, for which a simple geometrical parameters scanning is not
appropriate to provide a relevant optimization of the design.

\section{Acknowledgments}

The authors would like to thank St\'ephane Larouche for fruitful discussions and simulations concerning the
anti-reflective behavior of the grating and R\'emi Poll\`es for the help for figure \ref{resonance}.

\section{Appendix\label{appendix}}

The absorption inside the active a-Si:H layer is computed using either the scattering matrix method
\cite{krayzel10} for the flat coating  or a Fourier Modal Method\cite{lalanne,granet} for a
patterned electrode.  The absorption
is obtained by the difference between the flux of the Poynting vector
at the upper and a the lower interfaces of the silicon layer, when the
incident power is normalized to one. In TM polarization, this flux can
be written
\begin{equation}
\Phi_{M}=\Re \left(\sum_{n} \frac{1}{2\omega\epsilon_0\epsilon_r}
\gamma_n (A_n^*-B_n^*)(A_n+B_n)\right),
\end{equation}
where $A_n$ and $B_n$ are the amplitudes of mode $n$ in the silicon
layer (propagating or decreasing downward and upward respectively),
$\gamma_n=\sqrt{\epsilon_r\,k_0^2-\frac{2\pi\,n}{d}}$ its propagation
constant, $\epsilon_r$ being the permittivity of a-Si:H, $k_0$ the
wavenumber in vacuum and $d$ the period of the grating . In TE
polarization, the flux can be written
\begin{equation}
\Phi_{E}=\Re \left(\sum_{n} \frac{1}{2\omega\mu_0} \gamma_n^*
(A_n-B_n)(A_n^*+B_n^*)\right).
\end{equation}
For a flat structure, the flux is computed considering only the 0th
order ($n=0$) and there is no need to make any difference between the
two polarizations in normal incidence. The conversion efficiency is
averaged on TE and TM polarization when the absorption is different
for the two polarizations.

\bibliographystyle{els}

\end{document}